\documentclass[twocolumn,conference]{IEEEtran}
\usepackage[T1]{fontenc}
\usepackage[latin9]{inputenc}
\usepackage{float}
\usepackage{amsmath}
\usepackage{amssymb}
\usepackage{graphicx}
\usepackage[unicode=true,
 bookmarks=true,bookmarksnumbered=true,bookmarksopen=true,bookmarksopenlevel=1,
 breaklinks=false,pdfborder={0 0 0},pdfborderstyle={},backref=false,colorlinks=false]
 {hyperref}
\hypersetup{pdftitle={Your Title},
 pdfauthor={Your Name},
 pdfpagelayout=OneColumn, pdfnewwindow=true, pdfstartview=XYZ, plainpages=false}

\makeatletter

\providecommand{\tabularnewline}{\\}
\floatstyle{ruled}
\newfloat{algorithm}{tbp}{loa}
\providecommand{\algorithmname}{Algorithm}
\floatname{algorithm}{\protect\algorithmname}

\usepackage{algorithmic}
\floatname{algorithm}{Algorithm}
\usepackage{cite}
\usepackage{siunitx}
\usepackage{acronym}

\AtBeginDocument{\acrodef{AP}{access point}
\acrodef{CPU}{central processing unit}
\acrodef{CSI}{channel state information}
\acrodef{MIMO}{multiple\mbox{-}input multiple\mbox{-}output}
\acrodef{MISO}{multiple\mbox{-}input single\mbox{-}output}
\acrodef{MMSE}{minimum mean square error}
\acrodef{L-MMSE}{local minimum mean square error}
\acrodef{SNR}{signal-to-noise ratio}
\acrodef{SINR}{signal-to-interference plus noise ratio}
\acrodef{SE}{spectral efficiency}
\acrodef{EE}{energy efficiency}
\acrodef{SEmax}{spectral efficiency maximization}
\acrodef{EEmax}{energy efficiency maximization}
\acrodef{MR}{maximum ratio}
\acrodef{GP}{geometric programming}
\acrodef{NOMA}{non-orthogonal multiple access}
\acrodef{QoS}{quality of service}
\acrodef{MRC}{maximum-ratio combining}
\acrodef{APG}{accelerated projected gradient}
\acrodef{CDF}{cumulative distribution function}
\acrodef{AO}{alternating optimization}
}

\DeclareMathOperator{\maximize}{maximize}
\DeclareMathOperator{\minimize}{minimize}
\DeclareMathOperator{\st}{subject \, to}

\newcommand{\herm}{^{\mbox{\scriptsize H}}}
\newcommand{\trans}{^{\mbox{\scriptsize T}}}
\usepackage{xcolor}
\makeatother

\begin{document}
\title{{\color{black}A Low-Complexity Approach} for Max-Min Fairness in Uplink Cell-Free Massive
MIMO}
\author{\IEEEauthorblockN{Muhammad~Farooq\IEEEauthorrefmark{1},   Hien~Quoc~Ngo\IEEEauthorrefmark{2}, and Le~Nam~Tran\IEEEauthorrefmark{1}}\IEEEauthorblockA{\IEEEauthorrefmark{1}School of Electrical and Electronic Engineering,
University College Dublin, Ireland\\
 Email: muhammad.farooq@ucdconnect.ie; nam.tran@ucd.ie}\IEEEauthorblockA{\IEEEauthorrefmark{2}Institute of Electronics, Communications and
Information Technology, Queen's University Belfast, U.K.\\
 Email: hien.ngo@qub.ac.uk}}
\maketitle
\begin{abstract}
We consider  the problem of max-min fairness for uplink cell-free massive multiple-input multiple-output which is a potential technology for beyond 5G networks. More specifically, we aim to maximize the minimum spectral efficiency of all users subject to the per-user power constraint, assuming linear receive combining technique at access points. The considered problem can be further divided into two subproblems: the receiver filter coefficient design and the power control problem. While the receiver coefficient design turns out to be a generalized eigenvalue problem, and thus, admits a closed-form solution, the power control problem is numerically troublesome. To solve the power control problem, existing approaches rely on geometric programming (GP) which is not suitable for large-scale systems. To overcome the high-complexity issue of the GP method, we first reformulate the power control problem intro a convex program, and then apply a smoothing technique in combination with an accelerated projected gradient method to solve it. The simulation results demonstrate that the proposed solution can achieve almost the same objective but in much lesser time than the existing GP-based method. 
\end{abstract}


\section{Introduction}

\acresetall \bstctlcite{IEEEexample:BSTcontrol}Cell-free massive \ac{MIMO}
can be considered as the most recent development of distributed massive
\ac{MIMO} \cite{Ngo2017cfmm}, whereby a large number of \acp{AP}
serve a group of users in a large area of service. Cell-free massive
\ac{MIMO} has been receiving increasing attention as a potential
backbone technology for beyond 5G networks due to the massive performance
gains compared to the colocated massive \ac{MIMO} \cite{Ngo2017cfmm}.
While the research on cell-free massive \ac{MIMO} has been focusing on
the downlink, that is not the case for the uplink. For uplink cell-free
massive \ac{MIMO}, linear receivers are commonly employed at \acp{AP}
to combine users' signals. As a result, an important problem is to
jointly design linear combining and users' powers to maximize a performance
measure. In the following, we attempt to provide a comprehensive survey,
but by no means inclusive, on the previous studies on this line of
research. 

In \cite{Ngo2018rician}, the \ac{MRC} technique was employed at the
\acp{AP} based on local channel estimates to detect signals from single-antenna
users. Closed-form expression for achievable \ac{SE} is derived,
which is then used to formulate a max-min fairness problem. To solve
this problem, the work of \cite{Ngo2018rician} applies an \ac{AO} method by solving two subproblems: the linear combining
coefficient design which is in fact a generalized eigenvalue problem,
and the power control problem which is quasi-linear, and thus, can be solved by a bisection method with linear programming. A similar problem
is considered in \cite{Mai2019uplinkMU}, except the zero-forcing scheme
was used at \acp{AP}, and a similar solution to \cite{Ngo2018rician} is
proposed. In \cite{chamalee2019alternating}, the \ac{L-MMSE} combining
is adopted at each \ac{AP}. A similar \ac{AO} method was presented to
solve the max-min fairness problem. In particular, the power control
subproblem (when the \ac{L-MMSE} coefficients are fixed) is approximated
as a \ac{GP} problem. In \cite{Bashar2018mixQoS}, the problem of
max-min fairness for a group of users is considered, while imposing
\ac{QoS} constraints on the remaining users. This problem is also
solved using a \ac{GP} approach, which avoids the need of a bisection search as done in \cite{Ngo2018rician}. Other studies in the existing literature
such as \cite{Bashar2018maxmin,Bashar2019maxmin} also exploit \ac{GP}
to solve the power control problem. This is also the case in \cite{Bashar2019EEuplink}
where the problem of the total \ac{EE} maximization, subject to the per-user power and per-user \ac{QoS} constraints, is considered.
The problem of \ac{SE} maximization for cell-free massive \ac{MIMO}
with \ac{NOMA} is in studied \cite{Zhang2019noma}. Again, the considered
problem is solved using the sequential successive convex approximation method in combination
with \ac{GP}.

The main feature of all the above-mentioned studies is that they need to solve a single
geometric program or a series of linear programs to solve the incurred power control problems. It is well
known that these second-order methods requires very high complexity, and thus, are not
preferred for large-scale problems. This is why the existing literature
for the uplink of cell-free massive \ac{MIMO} has been limited to small-scale
scenarios whereby the number of \acp{AP} is around a few hundreds
(less than 250 in all aforementioned papers, to be precise). Thus,
the potential of cell-free massive \ac{MIMO} in large-scale settings has
remained unknown. We aim to fill this gap of the current literature
in this paper.

We assume \ac{MRC} at \acp{AP} using local channel estimates
and consider the problem of max-min fairness subject to power constraint
at each user. Similar to the known approaches, the considered problem
is split into two optimization subproblems. While the linear receiver
coefficient design is a generalized eigenvalue problem which admits
closed-form solution as in previous studies, we propose a more scalable
numerical method that can reduce the run-time to solve the power control
problem significantly. In this regard, our contributions are summarized
as follows:
\begin{itemize}
\item We \emph{reformulate} the power control problem, which is \emph{nonconvex}
in the original form, into a convex program. Due to the nature of
the max-min optimization, the objective of the equivalent convex program
is \emph{nonsmooth}, which can pose some numerical challenges.
\item We propose the use of Nesterov\textquoteright s smoothing technique
\cite{Nesterov2005a} to approximate the objective of the equivalent
convex function. This allows us to apply an \ac{APG} method that solves the approximated problem very fast. In
particular, our proposed method only requires the computation of the
gradient of the objective and closed-form expressions for projections.
\item We avail of the proposed method to explore the performance gains of
cell-free massive \ac{MIMO} in large-scale settings. In particular, we find
that having more \acp{AP} may be more beneficial than more antennas per \ac{AP}, given the number of the total antennas is kept the same.
\end{itemize}
\emph{Notations}: Bold lower and upper case letters represent vectors
and matrices. $\mathcal{CN}(0,a)$ denotes a complex Gaussian random
variable with zero mean and variance $a$. $\mathbf{X}\trans$ and
$\mathbf{X}\herm$ stand for the transpose and Hermitian of $\mathbf{X}$,
respectively. $x_{i}$ is the $i$-th entry of vector $\mathbf{x}$;
$[\mathbf{X}]_{i,j}$ is the $(i,j)$-th entry of $\mathbf{X}$. Notation
$\mathbf{e}_{i}$ denotes the $i$-th unit vector, i.e., the
vector such that $e_{i}=1$ and $e_{j}=0,\forall j\neq i$. $\nabla f(\mathbf{x})$
represents the gradient of $f(\mathbf{x})$. $\Vert\cdot\Vert$ denotes
the Euclidean or $l_{2}$ norm; $|\cdot|$ is the absolute value of the argument.

\section{System Model and Problem Formulation}

We consider an uplink cell-free massive \ac{MIMO} scenario where $K$ users
are jointly served by $M$ \acp{AP} and each \ac{AP} is equipped
with $L$ antennas. The users and \acp{AP} are randomly distributed
in a coverage area. We model the channel
coefficients between the $m$-th \ac{AP} and the $k$-th user as
$\mathbf{h}_{mk}=\beta_{mk}^{1/2}\mathbf{g}_{mk}$, where $\beta_{mk}$
is large-scale fading coefficient and $\mathbf{g}_{mk}\in\mathbb{C}^{L\times1}\sim\mathcal{CN}(0,1)$
is the vector of small-scale fading coefficients for all antennas
at the $m$-th \ac{AP}.

\subsection{Uplink Training and Channel Estimation}

We denote the length of the coherence interval and the uplink training
phase in data symbols as $T_{c}$ and  $T_{p}$, respectively. By $\sqrt{T_{p}}\boldsymbol{\psi}_{k}\in\mathbb{C}^{T_{p}\times1}$,
we denote the pilot sequence transmitted from the $k$-th user to
the \ac{AP}, where $\Vert\boldsymbol{\psi}_{k}\Vert^{2}=1$. For $\zeta_{p}$
being the transmit power of each pilot, the $m$-th \ac{AP} receives the
$L\times T_{p}$ matrix given as 
\begin{equation}
\mathbf{Y}_{m}^{p}=\sqrt{\zeta_{p}T_{p}}\sum_{k=1}^{K}\mathbf{h}_{mk}\boldsymbol{\psi}_{k}\herm+\mathbf{N}_{m}^{p},\label{eq:pilotMatrix}
\end{equation}
where $\mathbf{N}_{m}^{p}\in\mathbb{C}^{N\times T_{p}}\sim\mathcal{CN}(0,1)$
is the noise matrix during pilot transmission. Note that we have normalized
the channel coefficients $\mathbf{h}_{mk}$ by the square root of
the true noise power, $\sigma_{N}$, and thus the noise power in \eqref{eq:pilotMatrix}
is unity. To estimate the channel estimate at each \ac{AP}, the matrix
in \eqref{eq:pilotMatrix} is first multiplied with $\boldsymbol{\psi}_{k}$
to get $\mathbf{y}_{mk}^{p}=\mathbf{Y}_{m}^{p}\boldsymbol{\psi}_{k}$,
and then finding the minimum mean-square error (MMSE) estimate as
\cite{Ngo2017cfmm}
\begin{align*}
\hat{\mathbf{h}}_{mk} & =\mathbb{E}\{\mathbf{h}_{mk}\bigl(\mathbf{y}_{mk}^{p}\bigr)\herm\}\Bigl(\mathbb{E}\{\mathbf{y}_{mk}^{p}\bigl(\mathbf{y}_{mk}^{p}\bigr)\herm\}\Bigr)^{-1}\mathbf{y}_{mk}^{p}\\
 & =c_{mk}\mathbf{y}_{mk}^{p},
\end{align*}
where 
\begin{equation}
c_{mk}=\frac{\sqrt{\zeta_{p}T_{p}}\beta_{mk}}{\zeta_{p}T_{p}\sum_{i=1}^{K}\beta_{mi}\left|\boldsymbol{\psi}_{k}\herm\boldsymbol{\psi}_{i}\right|^{2}+1}.
\end{equation}
The mean square of any element of $\hat{\mathbf{h}}_{mk}$ is given
by
\begin{equation}
\nu_{mk}=\mathbb{E}\{\left|[\hat{\mathbf{h}}_{mk}]_{n}\right|^{2}\}=\frac{\zeta_{p}T_{p}\beta_{mk}^{2}}{\zeta_{p}T_{p}\sum_{i=1}^{K}\beta_{mi}\left|\boldsymbol{\psi}_{i}\herm\boldsymbol{\psi}_{k}\right|^{2}+1}.
\end{equation}

\subsection{Uplink Payload Data Transmission}

All the users simultaneously transmit payload data to the $m$-th
\ac{AP} during the payload interval $T_{c}-T_{p}$. (Here the downlink transmission is ignored.) We denote by
$s_{k}$ the symbol to be transmitted from the $k$-th user where
$\mathbb{E}\{|s_{k}|^{2}\}=1$. The transmitted signal from the $k$-th
user is $x_{k}=\sqrt{\eta_{k}}s_{k}$, where $\eta_{k}=\mathbb{E}\{|x_{k}|^{2}\}$
is the power at the $k$-th user. The received signal vector from
all users at the $m$-th \ac{AP} is expressed as
\begin{equation}
\mathbf{y}_{m}^{u}=\sum_{k=1}^{K}\mathbf{h}_{mk}\sqrt{\eta_{k}}s_{k}+\mathbf{n}_{m}^{u},
\end{equation}
where $\mathbf{n}_{m}^{u}\in\mathbb{C}^{L\times1}\sim\mathcal{CN}(0,1)$ is the noise at the $m$-th \ac{AP} during the uplink transmission.
The power at user $k$ must satisfy the following power constraint
\begin{equation}
\eta_{k}\leq\eta^{(\max)},\,\forall k,
\end{equation}
where $\eta^{(\max)}$ is the maximum transmit power. In this paper,
similar to \cite{Bashar2019EEuplink,Bashar2019maxmin}, \ac{MRC} is used at each \ac{AP}. For further improve the
performance of the signal detection at the \ac{CPU}, each \ac{AP} $m$ multiples
the resulting signal with a weighting coefficient $u_{mk}$ for the
$k$-th user. As a result, the aggregated received signal at the \ac{CPU}
is given by
\begin{equation}
r_{k}=\sum_{m=1}^{M}u_{mk}\hat{\mathbf{h}}_{mk}\herm\mathbf{y}_{m}^{u}.\label{eq:combinedsignal}
\end{equation}

\subsection{Achievable Uplink Spectral Efficiency}
To detect $s_{k}$, the combined signal at the \ac{CPU} in \eqref{eq:combinedsignal}
is split into desired signal, beamforming uncertainty,
inter-user interference, and total noise
for the $k$-th user. From this step, it was shown in \cite{Ngo2017cfmm}
that the following spectral efficiency for the $k$-th user is achievable
\begin{equation}
\mathcal{S}_{ek}(\boldsymbol{\eta},\mathbf{u})=\left(1-\frac{T_{p}}{T_{c}}\right)\log_{2}\bigl(1+\gamma_{k}(\boldsymbol{\eta},\mathbf{u})\bigr),\label{eq:SEk}
\end{equation}
where $\gamma_{k}(\boldsymbol{\eta},\mathbf{u})$ is the corresponding \ac{SINR}
for the $k$-th user defined as
\begin{equation}
\gamma_{k}(\boldsymbol{\eta},\mathbf{u})=\frac{\mathbf{u}_{k}\herm\bigl(\boldsymbol{\nu}_{kk}\boldsymbol{\nu}_{kk}\herm\eta_{k}\bigr)\mathbf{u}_{k}}{\mathbf{u}_{k}\herm\bigl(\sum_{i\neq k}^{K}\boldsymbol{\nu}_{ki}\boldsymbol{\nu}_{ki}\herm\eta_{i}+\frac{1}{L}\sum_{i=1}^{K}\mathbf{D}_{ki}\eta_{i}+\frac{1}{L}\mathbf{R}_{k}\bigr)\mathbf{u}_{k}},
\end{equation}
where $\boldsymbol{\eta}=[\eta_{1};\eta_{2};\ldots;\eta_{K}]$, $\mathbf{u}_{k}=[u_{1k};u_{2k};\ldots;u_{Mk}]$,
$\mathbf{u}=[\mathbf{u}_{1};\mathbf{u}_{2};\ldots;\mathbf{u}_{K}$],
$\mathbf{D}_{ki}\in\mathbb{R}_{+}^{M\times M}$ is a diagonal matrix
with $[\mathbf{D}_{ki}]_{m,m}=\nu_{mk}\beta_{mi}$, $\mathbf{R}_{k}\in\mathbb{R}_{+}^{M\times M}$
is a diagonal matrix with $[\mathbf{R}_{k}]_{m,m}=\nu_{mk}$, and
\begin{equation}
\boldsymbol{\nu}_{ki}\triangleq\left|\boldsymbol{\psi}_{k}\herm\boldsymbol{\psi}_{i}\right|\left[\nu_{1k}\frac{\beta_{1i}}{\beta_{1k}};\nu_{2k}\frac{\beta_{2i}}{\beta_{2k}};\ldots;\nu_{Mk}\frac{\beta_{Mi}}{\beta_{Mk}}\right].
\end{equation}

\subsection{Problem Formulation}

To deliver fairness of all users in the system, we consider the maximization
of the minimum \ac{SE} problem, which is mathematically stated as
\begin{equation}
(\mathcal{P}_{1}):\begin{cases}
\underset{\mathbf{u}_{k},\boldsymbol{\eta}}{\maximize} & \underset{1\leq k\leq K}{\min}\mathcal{S}_{ek}(\boldsymbol{\eta},\mathbf{u})\\
\st & \Vert\mathbf{u}_{k}\Vert=1,\,\forall k,\\
 & 0\leq\boldsymbol{\eta}\leq\eta^{(\max)}/L.
\end{cases}\label{eq:maxminProb}
\end{equation}
It is easy to see that the above problem is equivalent to maximizing
the minimum of the \ac{SINR} of individual users, which is given
by

\begin{equation}
(\mathcal{P}_{2}):\begin{cases}
\underset{\mathbf{u}_{k},\boldsymbol{\eta}}{\maximize} & \underset{1\leq k\leq K}{\min}\gamma_{k}(\boldsymbol{\eta},\mathbf{u})\\
\st & \Vert\mathbf{u}_{k}\Vert=1,\,\forall k,\\
 & 0\leq\boldsymbol{\eta}\leq\eta^{(\max)}/L.
\end{cases}\label{eq:equivProb}
\end{equation}
Due to the coupling of $\mathbf{u}_{k}$ and $\boldsymbol{\eta}$
in problem \eqref{eq:maxminProb}, it is quite natural to decouple
problem $(\mathcal{P}_{1})$ into two subproblems: the receiver coefficient
design and the power control problem. We remark that this method has
been widely used in the previous studies. We detail these two problems
in the following section and, in particular, present our novel solution
for the power control problem.

\section{Proposed Solution}

\subsection{Receiver Coefficient Design}

For the receiver coefficient design, we fix the power allocation variable
$\boldsymbol{\eta}$ and solve the following problem to find $\mathbf{u}_{k},\,\forall k$:
\begin{equation}
(\mathcal{P}_{3}):\begin{cases}
\underset{\mathbf{u}_{k}}{\maximize} & \underset{1\leq k\leq K}{\min}\gamma_{k}(\boldsymbol{\eta},\mathbf{u})\\
\st & \Vert\mathbf{u}_{k}\Vert=1,\,\forall k.
\end{cases}\label{eq:RxCoeff}
\end{equation}
This problem can be independently solved for each user to maximize
the \ac{SINR} at each user \cite{Bashar2018mixQoS}. Let us define
$\mathbf{A}_{k}=\boldsymbol{\nu}_{kk}\boldsymbol{\nu}_{kk}\herm\eta_{k}$
and $\mathbf{B}_{k}=\sum_{i\neq k}^{K}\boldsymbol{\nu}_{ki}\boldsymbol{\nu}_{ki}\herm\eta_{i}+\frac{1}{L}\sum_{i=1}^{K}\mathbf{D}_{ki}\eta_{i}+\frac{1}{L}\mathbf{R}_{k}$.
Then $(\mathcal{P}_{3})$ is reduced to the following problem for
each $k$:
\begin{equation}
(\mathcal{P}_{4}):\begin{cases}
\underset{\mathbf{u}_{k}}{\maximize} & \frac{\mathbf{u}_{k}\herm\mathbf{A}_{k}\mathbf{u}_{k}}{\mathbf{u}_{k}\herm\mathbf{B}_{k}\mathbf{u}_{k}}\\
\st & \Vert\mathbf{u}_{k}\Vert=1.
\end{cases}\label{eq:RxCoeff2}
\end{equation}
We note that problem $(\mathcal{P}_{4})$ is a generalized eigenvalue
problem. According to \cite[Lemma B.10]{Bjornson2017book}, the solution
to $(\mathcal{P}_{4})$ is given by

\begin{equation}
\mathbf{u}_{k}^{*}=\frac{\tilde{\mathbf{u}}_{k}}{\bigl\Vert\tilde{\mathbf{u}}_{k}\bigr\Vert}\label{eq:optimalU}
\end{equation}
 where $\tilde{\mathbf{u}}_{k}=\sqrt{\eta_{k}}\mathbf{B}_{k}^{-1}\boldsymbol{\nu}_{kk}$.
We remark that the receiver coefficient design has been presented
in the previous studies known to us. Our contributions in the paper
are those in handling the power control problem when the receiver
coefficients are held fixed. We provide the details in the next subsection.

\subsection{Power Control Problem}

After updating the receiver coefficients $\mathbf{u}_{k},\,\forall k$,
we need to find the power coefficients to further improve the objective
of $(\mathcal{P}_{3})$, which leads to the following power control
problem:
\begin{equation}
(\mathcal{P}_{5}):\begin{cases}
\underset{\boldsymbol{\eta}}{\maximize} & \underset{1\leq k\leq K}{\min}\gamma_{k}(\boldsymbol{\eta},\mathbf{u})\\
\st & 0\leq\boldsymbol{\eta}\leq\eta^{(\max)}/L.
\end{cases}\label{eq:powerAlloc}
\end{equation}
To solve the above problem, existing methods reformulate it using
an epigraph form, written as
\begin{equation}
(\mathcal{P}_{6}):\begin{cases}
\underset{\boldsymbol{\eta},t}{\maximize} & t\\
\st & 0\leq\boldsymbol{\eta}\leq\eta^{(\max)}/L,\\
 & t\leq\gamma_{k}(\boldsymbol{\eta},\mathbf{u}),\,\forall k.
\end{cases}\label{eq:GPprob}
\end{equation}
There are \emph{two main drawbacks} of such a method. First, the number
of constraints in $(\mathcal{P}_{6})$ has been increased by $K$.
This is not preferred for large-scale problems, i.e. when $K$ is
relatively large. Second, problem $(\mathcal{P}_{6})$ is in fact
a GP which can be solved by off-the-shelf convex solvers. The major
issue of GP is that it requires very high complexity, compared to
other standard convex problems, and thus is not suitable for large
scale problems. To overcome this complexity issue in the current literature,
we propose a more numerical scalable method in what follows.

\subsubsection{Convex Reformulation}

To derive the proposed method, we first reformulate the nonconvex
objective function in $(\mathcal{P}_{5})$ as 
\begin{equation}
\maximize\ \underset{1\leq k\leq K}{\min}\frac{1}{\gamma_{k}^{-1}(\boldsymbol{\eta})}\Longleftrightarrow\minimize\ \underset{1\leq k\leq K}{\max}\gamma_{k}^{-1}(\boldsymbol{\eta}),
\end{equation}
where 
\begin{align}
\gamma_{k}^{-1}(\boldsymbol{\eta}) & =\frac{\mathbf{u}_{k}\herm\bigl(\sum_{i\neq k}^{K}\boldsymbol{\nu}_{ki}\boldsymbol{\nu}_{ki}\herm\eta_{i}+\frac{1}{L}\sum_{i=1}^{K}\mathbf{D}_{ki}\eta_{i}+\frac{1}{L}\mathbf{R}_{k}\bigr)\mathbf{u}_{k}}{\mathbf{u}_{k}\herm\bigl(\boldsymbol{\nu}_{kk}\boldsymbol{\nu}_{kk}\herm\eta_{k}\bigr)\mathbf{u}_{k}}\nonumber\\
 & =\eta_{k}^{-1}\left(\sum_{i\neq k}^{K}a_{ki}\eta_{i}+\sum_{i=1}^{K}b_{ki}\eta_{i}+c_{k}\right).\label{eq:posynomialObj}
\end{align}
In the above equation, $a_{ki}$, $b_{ki}$, and $c_{k}$
are defined as 
\begin{align}
a_{ki} & =\frac{\mathbf{u}_{k}\herm\boldsymbol{\nu}_{ki}\boldsymbol{\nu}_{ki}\herm\mathbf{u}_{k}}{\mathbf{u}_{k}\herm\boldsymbol{\nu}_{kk}\boldsymbol{\nu}_{kk}\herm\mathbf{u}_{k}},\ 
b_{ki} =\frac{(1/L)\mathbf{u}_{k}\herm\mathbf{D}_{ki}\mathbf{u}_{k}}{\mathbf{u}_{k}\herm\boldsymbol{\nu}_{kk}\boldsymbol{\nu}_{kk}\herm\mathbf{u}_{k}},\nonumber\\
c_{k} & =\frac{(1/L)\mathbf{u}_{k}\herm\mathbf{R}_{k}\mathbf{u}_{k}}{\mathbf{u}_{k}\herm\boldsymbol{\nu}_{kk}\boldsymbol{\nu}_{kk}\herm\mathbf{u}_{k}}.
\end{align}
 Note that we have simply written $\gamma_{k}^{-1}(\boldsymbol{\eta})$
instead of $\gamma_{k}^{-1}(\boldsymbol{\eta},\mathbf{u})$ since
$\mathbf{u}$ is fixed. Now we use a change of variables to convert
the posynominal in \eqref{eq:posynomialObj} into a convex function.
Specifically, we define $\theta_{i}=\log\eta_{i}$ or $\eta_{i}=e^{\theta_{i}}$
and reformulate \eqref{eq:posynomialObj} as 
\begin{align}
f_{k}(\boldsymbol{\theta}) & =\gamma_{k}^{-1}(\boldsymbol{\theta})=e^{-\theta_{k}}\Bigl(\sum_{i\neq k}^{K}a_{ki}e^{\theta_{i}}+\sum_{i=1}^{K}b_{ki}e^{\theta_{i}}+c_{k}\Bigr)\nonumber \\
  & =\sum_{i\neq k}^{K}a_{ki}e^{(\mathbf{e}_{i}-\mathbf{e}_{k})\trans\boldsymbol{\theta}}+\sum_{i=1}^{K}b_{ki}e^{(\mathbf{e}_{i}-\mathbf{e}_{k})\trans\boldsymbol{\theta}}+c_{k}e^{-\mathbf{e}_{k}\trans\boldsymbol{\theta}}.\label{eq:convexapprox}
\end{align}
Let $f(\boldsymbol{\theta})=\underset{1\leq k\leq K}{\max}f_{k}(\boldsymbol{\theta})$.
Then $(\mathcal{P}_{5})$ can be reformulated as 
\begin{equation}
(\mathcal{P}_{7}):\begin{cases}
\underset{\boldsymbol{\theta}}{\minimize} & f(\boldsymbol{\theta})=\underset{1\leq k\leq K}{\max}f_{k}(\boldsymbol{\theta})\\
\st & \boldsymbol{\theta}\leq\theta_{\max},
\end{cases}\tag{\ensuremath{\mathcal{P}_{7}}}\label{eq:convexpowercontrol}
\end{equation}
where $\theta^{(\max)}=\log(\eta^{\max}/L)$. As mentioned earlier,
we can consider an epigraph form of $(\mathcal{P}_{7})$, which can
be solved by convex solvers. But this will increase the complexity.
A numerical difficulty in solving $(\mathcal{P}_{7})$ is due to the
fact that $f(\boldsymbol{\theta})$ is $nonsmooth$. To overcome
this issue, we apply Nesterov's smoothing technique to approximate
the function $f(\boldsymbol{\theta})$ by the following log-sum-exp
function \cite{Nesterov2005a} 
\begin{align}
f(\boldsymbol{\theta};\tau) & =\frac{1}{\tau}\log\frac{1}{K}\sum\nolimits _{k=1}^{K}e^{\tau f_{k}(\boldsymbol{\theta})},\label{eq:smoothappx}
\end{align}
where $\tau>0$ is the positive \emph{smoothness} parameter. It was
shown in \cite{Nesterov2005a} that $f(\boldsymbol{\theta})+\frac{\log K}{\tau}\geq f(\boldsymbol{\theta};\tau)\geq f(\boldsymbol{\theta})$.
In other words, $f(\boldsymbol{\theta};\tau)$ is a differentiable
approximation of $f(\boldsymbol{\theta})$ with a numerical accuracy
of $\frac{\log K}{\tau}$. Thus, with a sufficiently high $\tau$,
$f(\boldsymbol{\theta})$ can be replaced by $f(\boldsymbol{\theta};\tau)$
and we can consider the following problem 
\begin{equation}
\boxed{(\mathcal{P}_{8}):\begin{cases}
\underset{\boldsymbol{\theta}}{\minimize} & f(\boldsymbol{\theta};\tau)\\
\st & \boldsymbol{\theta}\in\Theta\triangleq\{\boldsymbol{\theta}|\boldsymbol{\theta}\leq\theta_{\max}\}.
\end{cases}}\tag{\ensuremath{\mathcal{P}_{8}}}\label{eq:convexpowercontrol-1}
\end{equation}

\subsubsection{Accelerated Projected Gradient Method}

 We are now in a position to propose an efficient algorithm to solve
$(\mathcal{P}_{8})$, which is essentially an \ac{APG} \cite{Beck2009LIP}. The description of the proposed
algorithm is provided in Algorithm \ref{alg:mAPG}. Note that $\alpha>0$
is called the step size which should be sufficiently small to guarantee
its convergence. Also, the notation $P_{\Theta}(\mathbf{u})$ denotes
the projection onto $\Theta$, i.e., $P_{\Theta}(\mathbf{u})=\arg\min\bigl\{\Vert\mathbf{x}-\mathbf{u}\Vert\ |\ \mathbf{x}\in\Theta\bigr\}$.
\begin{algorithm}[th]
\caption{Proposed \ac{APG} Algorithm for Solving $(\mathcal{P}_{8})$}
\label{alg:mAPG}

\begin{algorithmic}[1]

\STATE Input: $\boldsymbol{\theta}^{0}>0,\alpha>0$

\STATE $\boldsymbol{\theta}^{1}=\boldsymbol{\theta}^{0}$; $t_{0}=t_{1}=1$;
$n=1$

\REPEAT

\STATE $\mathbf{y}^{n}=\boldsymbol{\theta}^{n}+\frac{t_{n-1}-1}{t_{n}}(\boldsymbol{\theta}^{n}-\boldsymbol{\theta}^{n-1})$

\STATE $\boldsymbol{\theta}^{n+1}=P_{\Theta}(\mathbf{y}^{n}-\alpha\nabla f(\mathbf{y}^{n}))$
\label{grady}

\STATE $t_{n+1}=0.5\left(1+\sqrt{4t_{n}^{2}+1}\right)$;

\STATE $n=n+1$

\UNTIL{convergence }

\STATE Output: $\boldsymbol{\theta}^{*}$

\end{algorithmic}
\end{algorithm}

There are two main operations of Algorithm \ref{alg:mAPG}, namely:
finding the gradient of $f(\boldsymbol{\theta};\tau)$, and the projection
onto the feasible set $\Theta$. The details of these two operations
are given in the following. 

\subsubsection*{Gradient of $f(\boldsymbol{\theta};\tau)$}

It is easy to see that the gradient of $f(\boldsymbol{\theta};\tau)$
is given by
\begin{equation}
\nabla f(\boldsymbol{\theta};\tau)=\frac{1}{\sum\nolimits _{k=1}^{K}\exp\bigl(\tau f_{k}(\boldsymbol{\theta})\bigr)}\sum\nolimits _{k=1}^{K}e^{\tau f_{k}(\boldsymbol{\theta})}\nabla f_{k}(\boldsymbol{\theta}),\label{eq:gradconvexfunction}
\end{equation}
where $\nabla f_{k}(\boldsymbol{\theta})$ is found as
\begin{align}
\nabla f_{k}(\boldsymbol{\theta}) & =\sum_{i\neq k}^{K}a_{ki}e^{(\mathbf{e}_{i}-\mathbf{e}_{k})\trans\boldsymbol{\theta}}\bigl(\mathbf{e}_{i}-\mathbf{e}_{k}\bigr)\nonumber \\
 & +\sum_{i=1}^{K}b_{ki}e^{(\mathbf{e}_{i}-\mathbf{e}_{k})\trans\boldsymbol{\theta}}\bigl(\mathbf{e}_{i}-\mathbf{e}_{k}\bigr)  -c_{k}e^{-\mathbf{e}_{k}\trans\boldsymbol{\theta}}\mathbf{e}_{k}.\label{eq:gradInvSINR}
\end{align}

\subsubsection*{Projection onto $\Theta$}

For a given $\mathbf{x}\in\mathbb{R}^{K\times1}$, $P_{\Theta}(\mathbf{x})$
is the solution to the following problem: 
\begin{multline}
\underset{\boldsymbol{\theta}\in\mathbb{R}^{K\times1}}{\minimize}\ \Bigl\{\Vert\boldsymbol{\theta}-\mathbf{x}\Vert^{2}\ \Bigl|\ \boldsymbol{\theta}\leq\theta^{(\max)}\Bigr\}.
\end{multline}
It is straightforward to check that the solution to the above problem
is $\boldsymbol{\theta}=[\theta_{1};\theta_{2};\ldots;\theta_{K}]$,
where
\begin{equation}
\theta_{k}=\begin{cases}
x_{k}, & x_{k}\leq\theta_{k}\\
\theta^{(\max)}, & x_{k}>\theta_{k}
\end{cases},k=1,\ldots,K.\label{eq:projPolyhedral}
\end{equation}

In summary, combining the receiver coefficient design and the power
control problem, the proposed algorithm for the considered problem
is outlined in Algorithm \ref{alg:minmax-APG}. 
\begin{algorithm}[th]
\caption{Proposed \ac{APG}-based Algorithm for Solving $(\mathcal{P}_{1})$}
\label{alg:minmax-APG}

\begin{algorithmic}[1]

\STATE Input: $\boldsymbol{\theta}^{0}>0$; $m=1$

\REPEAT

\STATE For fixed $\boldsymbol{\theta}^{m-1}$, find optimal $\mathbf{u}^{m}=[\mathbf{u}_{1}^{m};\mathbf{u}_{2}^{m};\ldots;\mathbf{u}_{K}^{m}]$
using \eqref{eq:optimalU}\label{line:rxcoeff}

\STATE For fixed $\mathbf{u}^{m}$, find $\boldsymbol{\theta}^{m}=\boldsymbol{\theta}^{*}$
using Algorithm \ref{alg:mAPG}

\STATE $m=m+1$

\UNTIL{convergence}

\STATE Output: $\boldsymbol{\theta}^{m},\mathbf{u}^{m}$

\end{algorithmic}
\end{algorithm}

\subsection{Complexity Analysis}

We now provide the complexity analysis of the proposed algorithm for
one iteration using the big-O notation. It is clear that the complexity
of Algorithm \ref{alg:minmax-APG} is dominated by the computation
of four quantities: the receiver coefficients $\mathbf{u}^{m}$, the
smooth approximate objective function $f(\boldsymbol{\theta};\tau)$,
the gradient of $f(\boldsymbol{\theta};\tau)$, and the projection
onto $\Theta$. The computation of each $\mathbf{u}_{k}$
requires the calculation of the inverse of $\mathbf{B}_{k}$. Normally
this step requires the complexity of $\mathcal{O}(M^{3})$. However,
by exploiting the specific structure of $\mathbf{B}_{k},$ in the Appendix
we show that the complexity of this step is reduced to $\mathcal{O}\bigl((K-1)M^{2}\bigr).$
This \emph{massive computation saving when $M$ is large} has not
been reported in the existing literature. Accordingly, the complexity
for updating the receiver coefficient $\mathbf{u}^{m}$ is $\mathcal{O}(K^{2}M^{2})$.
It is easy to see that $KM$ multiplications are required to compute
$\mathcal{S}_{ek}(\boldsymbol{\eta})$. Therefore, the complexity
of finding $f(\boldsymbol{\eta})$ is $\mathcal{O}(K^{2}M).$ Similarly,
we can find that the complexity of $\nabla f_{k}(\boldsymbol{\theta})$
is $\mathcal{O}(K^{2}M)$. The projection of $\boldsymbol{\eta}$
onto $\Theta$ is given in \eqref{eq:projPolyhedral}, and thus the
complexity of the projection is $\mathcal{O}(K)$. In summary, the
per-iteration complexity of the proposed algorithm for solving $(\mathcal{P}_{4})$
is $\mathcal{O}(K^{2}M^{2})$. 

\section{Numerical Results}

We evaluate the performance of our proposed method in terms of the
achievable rate and the time complexity. In the considered cell-free
massive \ac{MIMO} system, we randomly distribute the \acp{AP} and
the users over a $D\times D$ \si{\km\squared}. The large-scale fading
coefficient between the $m$-th AP and the $k$-th user is generated
as $\bar{\beta}_{mk}=\mathrm{PL}_{mk}.z_{mk},$ where $\mathrm{PL}_{mk}$
is the path loss between the $m$-th \ac{AP} and the $k$-th user,
and $z_{mk}$ represents the log-normal shadowing between the $m$-th
\ac{AP} and the $k$-th user with mean zero and standard deviation
$\sigma_{\textrm{sh}}$, respectively. In this paper, we adopt the
three-slope path loss model and model parameters as in \cite{Farooq2020}.
The noise power is given by $B\times k_{B}\times T_{0}\times W$,
where $B=20$ MHz denotes the bandwidth, $k_{B}=1.381\times10^{-23}$ (Joule/Kelvin)
represents the Boltzmann constant, $T_{0}=290$ (Kelvin) denotes the
noise temperature, and $W=9$ dB shows the noise figure. The length
of the coherence time and the uplink training phase are set to $T_{c}=200$, $T_{p}=20$, respectively. If not otherwise mentioned, we set the power transmit
power for downlink data transmission and uplink training phase (before
normalization) as $\zeta_{u}=0.2$ W and $\zeta_{p}=0.2$ W.

\begin{figure}[tbh]
\centering\includegraphics[bb=62bp 552bp 291bp 738bp]{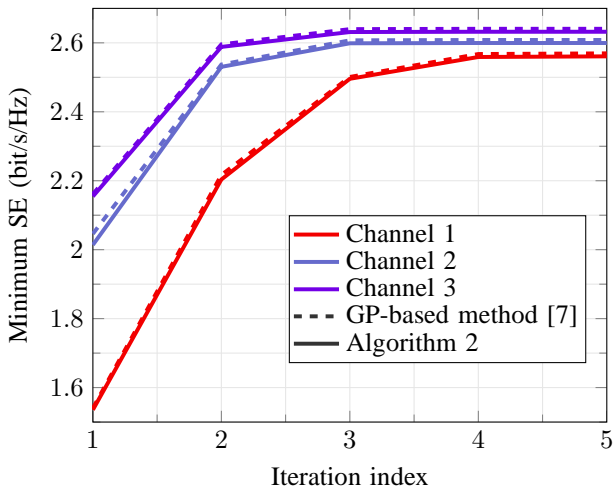} \caption{Convergence of Algorithm \ref{alg:minmax-APG} and the \ac{GP}-based
algorithm for three random channel realizations. The relevant parameters
are taken as $M=150,K=20,L=1,D=1$.}
\label{Fig:GPconvergence}
\end{figure}

In the first numerical experiment, we compare Algorithm \ref{alg:minmax-APG}
with the method in \cite{Bashar2018mixQoS,Bashar2019maxmin} which
solves \eqref{eq:maxminProb} using \ac{GP} for the power control
problem. We refer to this method as the GP-based method in this section.
For this method, we use the convex solver MOSEK \cite{MOSEKApS15}
through the modeling tool CVX \cite{cvx} in the GP mode to
solve the power control problem in $(\mathcal{P}_{1})$. Figure \ref{Fig:GPconvergence}
shows the convergence of Algorithm \ref{alg:minmax-APG} and the \ac{GP}-based
method to solve $(\mathcal{P}_{1})$ for three randomly generated
channel realizations.

We can see that both methods reach almost the same objective for all
three considered channel realizations. The marginal difference is
due to a moderate value of the smoothness parameter $\tau$ in \eqref{eq:smoothappx}.
The main advantage of our proposed method over the \ac{GP}-based
method is that each iteration of the proposed method is very memory
efficient and computationally cheap, and hence, can be executed very
fast. As a result, the total run-time of the proposed method is far
less than that of the \ac{GP}-based method as shown in Table~\ref{table: Table 1}.
In Table~\ref{table: Table 1}, we report the actual run-time of
both methods to solve the max-min problem. Here, we run our codes
on a 64-bit Windows operating system with 16 GB RAM and Intel CORE
i7, 3.7 GHz. Both iterative methods are terminated when the difference
of the objective for the latest two iterations is less than $10^{-5}$.
\begin{table}[tbh]
\caption{Comparison of run-time (in seconds) between Algorithm \ref{alg:minmax-APG}
and the \ac{GP}-based method. Here, $K=20$ and $D=1$.}
\label{table: Table 1} \centering{}%
\begin{tabular}{c|c|c}
\hline 
APs & \ac{GP}-based Method & Proposed Method\tabularnewline
\hline 
120 & 3.67 & \textbf{0.80}\tabularnewline
\hline 
160 & 3.79 & \textbf{1.21}\tabularnewline
\hline 
200 & 4.08 & \textbf{1.86}\tabularnewline
\hline 
240 & 4.42 & \textbf{2.45}\tabularnewline
\hline 
\end{tabular}
\end{table}

\begin{figure}[H]
\centering\includegraphics[bb=62bp 552bp 291bp 738bp]{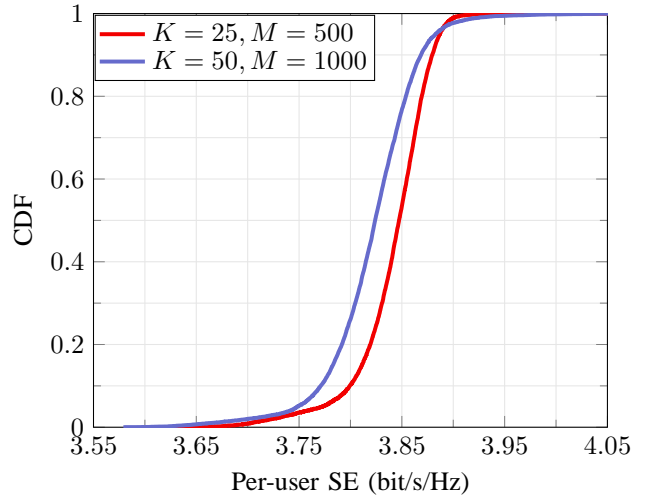}\caption{\ac{CDF} of per-user achievable \ac{SE} for two scenarios $K=25,M=500$
and $K=50,M=1000$. The number of antennas at each AP is $L=1$.}
\label{Fig:CDFplot}
\end{figure}

In the next experiment, we plot the \ac{CDF} of the per-user \ac{SE} (bit/s/Hz) for $300$ channel realizations shown in Fig. \ref{Fig:CDFplot}.
We consider two large-scale scenarios: (i) $K=25,M=500$ and (ii)
$K=50,M=1000$. Note that the ratio between the number of APs to the
number of users is the same for two cases. It can be seen in Fig.
\ref{Fig:CDFplot} that as the number of the users increases, the
per-use \ac{SE} slightly decreases. We also observe that the per-user \ac{SE} differs in the range of 0.1 bit/s/Hz which means that
the fairness is indeed achieved among the users. 

Finally, we investigate the effect of increasing the number of antennas
per \ac{AP} on the minimum achievable \ac{SE}. Specifically, we plot
the minimum achievable \ac{SE} with respect to the number of antennas
for both $M=250$ and $M=500$ \acp{AP}. The number of users is fixed to
$K=50$. As expected, the minimum achievable \ac{SE} increases with the
number of antennas per \ac{AP} but the increase tends to be small when
the number of antennas is sufficiently large. The reason is that for
a large number of \acp{AP} channel harderning and favorable propagation
can be achieved by a few antennas per \ac{AP}. Specifically, we can see
that the \ac{SE} for the case of $500$ \acp{AP} and $5$ antennas per \ac{AP}
is larger than the \ac{SE} for the case of 250 \acp{AP} and 10 antennas per \ac{AP}. Therefore, for large-scale cell-free massive MIMO having more
APs with a few antennas each tends to be more beneficial than having
fewer APs with more antennas each.
\begin{figure}[H]
\includegraphics[bb=62bp 550bp 294bp 738bp]{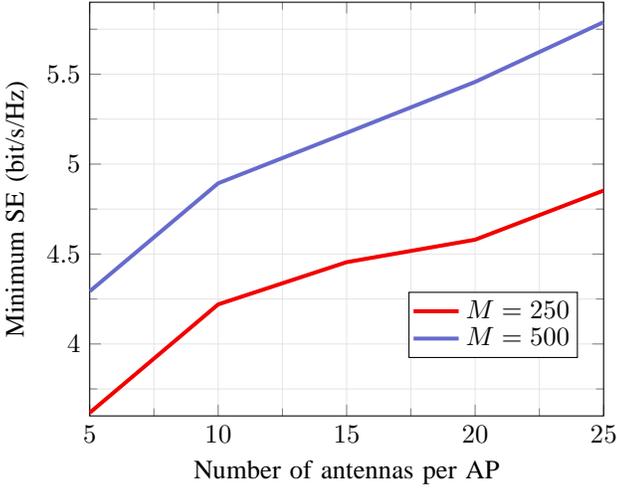}\caption{Maximized minimum achievable \ac{SE} with respect to the number of antenna
at each AP for $K=50,D=1.$}
\label{Fig:RatevsAntennas}
\end{figure}

\section{Conclusion}

In this paper, we have proposed a low-complexity method for maximizing
the minimum achievable \ac{SE} in the uplink of the cell free massive
\ac{MIMO} subject to per-user power constraint, assuming the \ac{MRC}
technique at \acp{AP}. As in previous studies, we have divided the min-max
problem into two subproblems: the receiver coefficient design and
the power control problem. For the power control problem, the existing
solutions rely on \ac{GP} which requires high complexity. To overcome
this issue, we have reformulated the power control problem into a
convex form and then proposed an efficient numerical algorithm based
on Nesterov's smoothing technique and \ac{APG}.
Our proposed solution only requires the first order information of
the objective and the projection, both of which are given in the closed-form.
We have numerically shown that our proposed solution practically achieves
the same objective as the \ac{GP}-based method but in much lesser
time. We have used the proposed method to study the performance of
the large-scale systems up to 1000 \acp{AP} for which the known methods
are not suitable. We have also shown that a large-scale cell-free
massive \ac{MIMO} having more \acp{AP} with a few antennas has better
performance than a similar system having a fewer \acp{AP} with more
antennas per \ac{AP}.

\appendix{}

In this appendix, we show that the complexity of computing $\mathbf{u}_{k}$
is $\mathcal{O}(KM^{2})$. Without loss of generality, let us consider
the inverse of $\mathbf{B}_{k}$ for user $1$. It is obvious that
we can write $\mathbf{B}_{1}^{-1}$ as
\begin{align}
\mathbf{B}_{1}^{-1} & =\Bigl(\sum_{i=2}^{K}\boldsymbol{\nu}_{1i}\boldsymbol{\nu}_{1i}\herm\eta_{i}+\tilde{\mathbf{D}}_{1}\Bigr)^{-1}\nonumber \\
 & =\tilde{\mathbf{D}}_{1}^{-1}-\frac{\tilde{\mathbf{D}}_{1}^{-1}\boldsymbol{\nu}_{12}\boldsymbol{\nu}_{12}\herm\tilde{\mathbf{D}}_{1}^{-1}}{1+\boldsymbol{\nu}_{12}\herm\mathbf{D}_{k}^{-1}\boldsymbol{\nu}_{12}}
\end{align}
where $\tilde{\mathbf{D}}_{1}=\frac{1}{L}\sum_{i=1}^{K}\mathbf{D}_{1i}\eta_{i}+\frac{1}{L}\mathbf{R}_{1}+\sum_{i=3}^{K}\boldsymbol{\nu}_{1i}\boldsymbol{\nu}_{1i}\herm\eta_{i}$.
If we further express $\tilde{\mathbf{D}}_{1}$ in terms of
the inverse of $\frac{1}{L}\sum_{i=1}^{K}\mathbf{D}_{1i}\eta_{i}+\frac{1}{L}\mathbf{R}_{1}+\sum_{i=4}^{K}\boldsymbol{\nu}_{1i}\boldsymbol{\nu}_{1i}\herm\eta_{i}$ and keep on repeating this step,
then finally we need to compute the inverse of the term $\frac{1}{L}\sum_{i=1}^{K}\mathbf{D}_{1i}\eta_{i}+\frac{1}{L}\mathbf{R}_{1}$
which is a diagonal matrix, and thus only requires $\mathcal{O}(M)$.
The complexity of each step in the above process is $\mathcal{O}(M^{2})$,
and thus the computation of $\mathbf{B}_{1}^{-1}$ is $\mathcal{O}\bigl((K-1)M^{2}\bigr)$.
Multiplying $\mathbf{B}_{1}^{-1}$ with $\boldsymbol{\nu}_{11}$ to
obtain $\mathbf{u}_{1}$ takes additional $\mathcal{O}(M)$, and thus,
the complexity of the computation of $\mathbf{u}_{1}$ stands at $\mathcal{O}\bigl(KM^{2}\bigr)$. 

\bibliographystyle{IEEEtran}
\bibliography{IEEEabrv,uplink_ref}

\end{document}